# Simulations of Gamma-ray emission from magnetized micro-quasar jets

Odysseas Kosmas[1][*] and Theodoros Smponias[2][†]
[1] *Modelling and Simulation Centre, MACE, University of Manchester, Sackville Street, Manchester, UK* and
[2] *Division of Theoretical Physics, University of Ioannina, GR-45110 Ioannina, Greece*

In this work, we simulate $\gamma$-rays created in the hadronic jets of the compact object in binary stellar systems known as microquasars. We utilize as main computational tool the 3-D relativistic magneto-hydro-dynamical code PLUTO combined with in house derived codes. Our simulated experiments refer to the SS433 X-ray binary, a stellar system in which hadronic jets have been observed. We examine two new model configurations that employ hadron-based emission mechanisms. The simulations aim to explore the dependence of the $\gamma$-ray emissions on the dynamical as well as the radiative properties of the jet (hydrodynamic parameters of the mass-flow density, gas-pressure, temperature of the ejected matter, high energy proton population inside the jet plasma, etc.). The results of the two new scenarios of initial conditions for the micro-quasar stellar system studied, are compared to those of previously considered scenarios.

[*] odysseas.kosmas@manchester.ac.uk
[†] t.smponias@hushmail.com

# I. INTRODUCTION

The emissions of $\gamma$-rays, neutrinos, etc. within the jets of microquasars (MQs) have recently gained great interest among the researchers seeking to understand the structure properties and evolution of X-ray binary systems [1–3].

Special interests appeared on the $\gamma$-ray emission mechanisms inside the hadronic jets, as the photon-hadron interactions [4, 5], the hadron-hadron interactions [6, 7] as well as the $\gamma$-ray absorption, that help to deepen our knowledge on microquasars evolution [8].

On the other hand, the strong magnetic field in the jets may significantly affect the total internal $\gamma$-ray and neutrino emissions by tuning several processes determining the high energy proton population (synchrotron radio emission, etc.). Therefore, magnetic field effects should be appropriately incorporated and treated in jet models [9, 10].

Recently, neutrinos from galactic microquasars, even though not being detected so far, have been modelled and several simulations have been performed towards this aim. Such modelling may support future attempts to detect them (see e.g. Ref. [7, 11]).

Invariably, the jets of micro-quasars as well as in general the astrophysical jets, may be described as fluid flow emanating from the vicinity of the compact object. Such a micro-quasar system is the SS433 X-ray binary consisted of a donor (companion) star and a compact stellar object which emits relativistic jets in various wavelength bands. Up to now, it is the only microquasar observed with a definite hadronic content in its jets, as verified from observations of spectral lines [1, 2, 9].

Radiative transfer calculations may be performed at every point in the jet (for a range of frequencies/energies, at every location) [12], providing the relevant emission and absorption coefficients. In such cases, finally a line-of-sight integration may derive synthetic images of jet $\gamma$-ray emission, at the energy-window of interest [12, 13].

The relativistic treatment of jets, takes into account various energy loss mechanisms that occur through several hadronic processes [4–7]. In the known fluid approximation, macroscopically the jet matter behaves as a fluid collimated by the magnetic field. At a smaller scale, consideration of the kinematics of the jet plasma becomes necessary for treating shock acceleration effects.

Many authors consider that, the proton-proton (p-p) collisions between fast high energy protons (non-thermal protons) and bulk flow slow (thermal) protons constitute the dominant cooling process of the high energy proton population of the jet. This mechanism explains, further, the main part of the $\gamma$-rays and neutrinos produced in the binary SS433 system. The acceleration of thermal protons (diffusive first-order shock acceleration) occurs above a minimum threshold proton energy [7].

Assuming a Maxwellian energy distribution for the 'slow' protons, only a tiny portion of the total bulk proton jet-flow, i.e. the fastest of them, may undergo diffusive shock acceleration and may jump to the fast proton population. Hence, the fast protons constitute a small fraction of the total jet proton density which subsequently produce $\gamma$-rays, neutrinos, etc. In this work, we assume that this is the dominant mechanism generating high-energy $\gamma$-rays in the SS433 microquasar jets.

For the sake of completeness, we mention that another rather important mechanism has been suggested based on the hadronic interactions occurring within the jet-wind interaction zone [7]. In this scenario, the $\gamma$-rays are generated from the decay of neutral pions, as $\pi^0 \rightarrow \gamma + \gamma$. Pions are created via inelastic collisions of jet protons, ejected from the compact object, and ions of the stellar wind (such a process may also occur in the vicinity of the extended disk of the binary system) [6]. The latter emission mechanism is rather weak in SS433 [7].

So far, microquasar $\gamma$-ray emissions have been observed through Cerenkov telescopes (HESS, MAGIC, CTA) and orbital telescopes (INTEGRAL, Fermi) [14–19]. We also mention that, for low energy $\gamma$-rays, ongoing and future or next generation measurements with INTEGRAL (ESA satellite) and Fermi (NASA orbital telescope) may provide new data. Furthermore, very high energy $\gamma$-rays, in general above about $30 GeV$, can be studied with ground based Cerenkov telescopes [20].

Phenomenologically, estimations of high energy $\gamma$-ray emission from microquasars have extensively been carried out [9, 12]. In this work, using the 3-D relativistic hydrocode PLUTO [21] and some in house (mainly radiative transfer) codes (now written both in Mathematica and in C) [22–24], we model $\gamma$-ray emissions from hadronic microquasar jets in the $E_\gamma$-energy range $1.2 GeV \leq E_\gamma \leq 10^2 - 10^3$ TeV.

The emission/absorption coefficients are computed on the basis of Monte Carlo simulations of terrestrial particle-particle collisions experimental data [12, 25, 26] that describe $\gamma$-ray emission in MQs. Such simulations provide analytical parametrizations for emission and absorption coefficients in a wide range of $\gamma$-ray energies (frequencies) produced in microquasar jets [12, 27].

Furthermore, by exploiting the hydrodynamic variable values supplied by PLUTO, our line of sight code may provide emission/absorption coefficients for every location in the jet. The results produced this way depend on the initial high-energy proton-distribution inserted in the hydrodynamical model jet [12, 28].

In the rest of the paper, at first (Sect. II) the main MQs emissions mechanisms are briefly summarized. Then (Sect. II D), the radiative transfer method and the calculational procedure for obtaining gamma-ray emission, is briefly described. The results of the 3-D relativistic hydrocode PLUTO for the emission/absorption coefficients are presented in Sect. IV. Finally (Sect. V), the main conclusions extracted in this work are summarized.

| Parameter/Scenario | C (run3) | D (run4) | Comments |
|---|---|---|---|
| cell size ($\times 10^{10} cm$) | 0.40 | 0.40 | PLUTO's computational cell |
| $\rho_{jet}$ ($cm^{-3}$) | $1.0 \times 10^{11}$ | $1.0 \times 10^{14}$ | jet's matter density |
| $\rho_{sw}$ ($cm^{-3}$) | $1.0 \times 10^{11}$ | $1.0 \times 10^{12}$ | stellar wind density |
| $\rho_{adw}$ ($cm^{-3}$) | $1.0 \times 10^{11}$ | $1.0 \times 10^{13}$ | accretion disk wind density |
| $t_{run}^{max}$ (s) | $1.5 \times 10^3$ | $1.5 \times 10^3$ | model execution time |
| Method | P. L. | P. L. | Piecewise Linear |
| Integrator | Ch. Tr. | Ch. Tr. | Characteristic Tracing |
| EOS | Ideal | Ideal | Equation of state |
| n | 0.1005 | 0.1005 | $E_\gamma$=100 GeV normalisation |
| BinSep (cm) | $4.0 \times 10^{12}$ | $4.0 \times 10^{12}$ | Binary star separation |
| $M_{BH}/M_\odot$ | 3-10 | 3-10 | Mass range of collapsed star |
| $M_\star/M_\odot$ | 10-30 | 10-30 | Mass range of Main Seq. star |
| $\beta = v_0/c$ | 0.26 | 0.26 | Initial jet speed |
| $L_k^p$ | $10^{36}$ | $10^{39}$ | Jet kinetic luminocity |
| grid resolution | 300*500*300 | 300*500*300 | PLUTO grid resolution (xyz) |

TABLE I. Scenario C (run 3) has artificially accelerated precession, while scenario D (run 4) has all the densities of the system increased by a few orders of magnitude, in order to account for a higher jet-mass flow-rate (jet's kinetic luminocity). The parameter n refers to a 'normalization' process that equates the results of two different methods of $\gamma$-ray emission calculations, one applied for energies above $E_\gamma$=100 GeV, and the other below this limit.

## II. OUTLINE OF MQS JET EMISSION MECHANISMS

The SS433 microquasar, an eclipsing X-ray binary system with a compact object most likely a black hole, comprises two oppositely directed precessing hadronic jets. The spectrum of the companion (donor) star suggests that it is rather a late A-type micro-quasar. In modelling $\gamma$-ray emission from SS433 in our present work, we assume that they are created mainly through p-p interactions between fast (relativistic) and slow (cold) protons within its hadronic jets.

Other production mechanisms, though not excluded, are considered less important. For example, some authors considered that the high-energy $\gamma$-rays in hadronic MQs jets, are produced from p-p collisions taking place in the jets due to the interaction of relativistic protons with target protons of the rather weak stellar wind created in the companion star [6].

The main reaction chain that produces $\gamma$-rays starting from p-p interaction through the pion decay channel, is written as

$$p + p \rightarrow p + p + \pi^0 \rightarrow p + p + 2\gamma. \qquad (1)$$

($m_p = 1.67 * 10^{-24}$ g and $m_\pi = 2.38 \times 10^{-25}$ g). In Refs. [6, 7, 25] an analytical description of the evolution of reaction chain within the jet is presented. Here we assume that a very energetic but small proton population, $N_{fp}$, (formed due to shock fronts in the jet), interacts with the bulk flow jet protons.

From the latter protons, high energy protons are produced through first order Fermi acceleration that occurs at shocks within the jet [7]. Such shocks are considered rather homogeneously distributed throughout the jet. The jet matter density is closely related to the density of the aforementioned shocks, thus, the internal shocks convert a portion of the bulk kinetic energy of cold protons, K, to the fast protons energy $E_p$ of the multi-directional motion. The rate, $t_{acc}^{-1}$, at which some slow protons are transferred to the high-energy distribution is described by [1]

$$r = t_{acc}^{-1} = E^{-1}\frac{dE}{dt} \simeq \beta^2 \frac{ceB}{E_p}, \qquad (2)$$

where $e$ denotes the proton charge, $\beta = u_{jet}/c$ with $u_{jet}$ being the jet matter's local velocity, and $B$ denotes the magnetic field.

Concerning the magnetic field $B$, we assume that this is either constant or it decreases with the distance from the jet base as $B \sim z^{-1}$ [29, 30]. We stress that, such a variation leads to a decrease of proton acceleration not more than two orders of magnitude compared to its value around the jet's base $z = z_0$. We stress that, the acceleration rate of fast protons, $r = t_{acc}^{-1}$, depends on the magnetic field $B$ as indicated in Eq. (2).

Alternatively, Eq. (2) gives the production rate of fast protons at every 'location' in the jet, though the production of $\gamma$-rays from these fast protons occurs at a next stage (by 'location' we mean a hydrodynamical grid-cell which, microscopically, is very large (of the order of $10^{10}$ cm) [7]. We note that, the presence of $\beta^2$ in Eq. (2), cuts off $\gamma$-ray emission from slow moving matter into the jet (acceleration sites are much less in slow matter).

Moreover, in jet emission calculations the $\beta^2$ is incorporated into the jet density. Also, the proton acceleration rate cannot affect the $\gamma$-ray emission rate, unless $\beta$ drops below some value. As we will see below, in this work, instead of the proton density $\rho$, we also use the product $\rho u^2$ [7, 13].

Regarding the particles ejected from a hadronic jet, we consider that they are mostly slow (thermal) protons of density $n_{sp}$ and a small portion of fast (non-thermal) protons of density $n_{fp}$. The energy distribution of the fast protons (in the jet's frame, see below) is described by [7, 13]

$$n_{fp}(E') = K_0 \left(E'\right)^{-\alpha} \tag{3}$$

$(E' \equiv E'_{fp})$ which is a power law type distribution. The parameter $\alpha$ takes the value $\alpha = 2$ and $K_0$ denotes a normalization constant [13].

In Table I, we tabulate the values of some model jet parameters (together with explanation of their symbols) relevant to $\gamma$-ray emissions from the SS433 binary system (The scenarios C and D are described below).

### A. Flux in observation and jet's frame

In our $\gamma$-ray flux calculations, we denote the flux density of the fast (slow) proton populations as $J_{fp}$ ($J_{sp}$), in the observation frame, and as $J'_{fp}$ ($J'_{sp}$), in the jet's frame. Moreover, at a given jet point, in the jet (moving) frame, the fast protons energy-spectrum is described by Eq. (3). For the corresponding fast-protons spatial density we adopt the relation

$$n_{fp}(E') \propto w \frac{dN'_{fp}}{dE'} = wK_0(E')^{-\alpha} \tag{4}$$

where

$$w = n_{sp}\beta^2 \tag{5}$$

($\beta \equiv u/c$, with $u = |\mathbf{u}|$ being the magnitude of the total velocity vector) where n$_{sp}$ is in protons/cm$^3$ [28]. From Eq. (5) one can obviously conclude that, the emission from the fast-moving matter of the jet, is larger compared to the emission from slow-moving matter which is because in Eq. (2), $t_{acc}$ is proportional to $\beta^2$ and creates fast proton jet-density which subsequently allows for p-p collisions to occur and for $\gamma$-rays to be produced. [6].

### B. The model of jet's dynamics

The jet is assumed to travel along the $y$ axis (we consider particles of mass $m_p$). Then, for the flux densities, we can write (steady state)

$$J_{sp} = m_p w = m_p n_{sp} \beta^2 \tag{6}$$

In general, $u$ is not necessarily parallel to the $y$-axis, but it may point almost anywhere which in turn means that, emission may occur from jet matter moving in any direction. Furthermore, the emission mechanism is based on "randomly oriented turbulent shocks", so the emission is considered multi-directional (no secondary emissions from scattering are assumed, since more shocks exist wherever the jet matter moves faster).

In our simulations, large turbulences of the jet flow may appear which favor shocks existence. This is due to the assumed strong dependence on the local velocity of the jet or ambient matter (acceleration rate is proportional to $u^2$ and further $J = \rho u^2$). Here, instead of the simple $\rho$ dependence, we adopt, in addition, the $\rho u^2$ dependence to distinguish the moving matter of the jet from that of the surrounding medium. This way, the calculation of $\gamma$-rays and neutrino emissions from the jet are de-coupled from the influence of the surrounding matter. Then, the jet's contribution to high energy γ-ray emission is mostly dependent on its internal turbulence (turbulence here means spatial number density of proton accelerating shocks randomly oriented) [7].

From the above discussion, we note in short that, the proton acceleration efficiency of the model jet is proportional to the square of the local velocity of the flow (the fast proton density is considered proportional to the square of the local velocity). Thus, the fast protons spatial density, $n_{fp}$, is also taken as proportional to the slow protons spatial density, $n_{sp}$ as well as to the square of the local velocity.

Furthermore, for hydrodynamical jets [12, 13] the fast proton current density, $J'_{fp}(E')$, as a function of their energy, is given by

$$J'_{fp}(E') = \frac{c}{4\pi} K_1 n_j \beta_j^2 (E')^{-\alpha} \tag{7}$$

In the latter expression, $n_j$ denotes the slow bulk jet protons local density hydrodynamical model (PLUTO code) [12].

### C. The current density at observer's frame

Regarding the transformation, to the observer frame, we write [31]

$$J_{fp}(E_p, t) = \frac{c}{4\pi} K_1 n_{sp} \beta^2 F . \tag{8}$$

$F$ represents a function of stationary frame energy $E_p$ written as

$$F = \frac{\gamma^{-\alpha+1}(E_p - \beta_b \sqrt{E_p^2 - m_p^2 c^4} \cos i_j)^{-\alpha}}{[\sin i_j{}^2 + \gamma^2(\cos i_j - (\beta_b E_p)/\sqrt{E_p^2 - m_p^2 c^4})^2]^{1/2}} \tag{9}$$

In the latter equation, $i_j(t)$ denotes the angle between the jet axis and the line of sight (for SS433 microquasar $\beta_b = v_b/c = 0.26$), and

$$\gamma = \left[1 - \beta_b^2\right]^{-1/2} , \tag{10}$$

is the jet Lorentz factor.

Thus, $F$ provides the relation of $J_{fp}$ (for laboratory frame) that depends on the γ-ray energy $E_\gamma$ as measured in laboratory frame (see Eq. 7). In conclusion, one can work with laboratory frame quantities only, which are also the jet model quantities (for other symbols the reader is referred to Ref. [12, 13, 27, 28]

### D. The 3-D radiative transfer in time-dependent jet

The propagation of γ-rays along a one-dimensional line-of-sight (without scattering) we address here, is based on the relation

$$\frac{dI_\nu}{dl} = -I_\nu \kappa_\nu + \epsilon_\nu \tag{11}$$

where $\kappa_\nu$ is the absorption coefficient at a given frequency (energy) $\nu$ and $\epsilon_\nu$ is the relevant emission coefficient. $I$ denotes the intensity and $l$ is the length along the line-of-sight.

By considering the model jet artificially imaged in γ-rays, we calculate the emission from a small jet element corresponding to a computational cell (for simulated emissions). To this aim, we first define the quantity

$$dI_\gamma = J_\gamma dV = \rho dV \frac{dN_\gamma}{dE_\gamma} = dm_{cell} \frac{dN_\gamma}{dE_\gamma} \tag{12}$$

to represent the intensity created from a cell of volume $dV$, at a given frequency, or γ-ray energy $E_\gamma$, while $\rho$ is the hydro density of the cell ($\rho = m_p n_{sp}$) and $dN_\gamma$ stands for the emission coefficient of the cell at the same frequency. An alternative version of the above quantity is

$$dI_\gamma = \rho dV \frac{dN_\gamma}{dE_\gamma} u^2 , \tag{13}$$

where $u$ is the local jet matter velocity.

### III. USE OF PLUTO CODE FOR GAMMA-RAY EMISSION CALCULATIONS

Our calculation of the emission coefficient $dN_\gamma$, proceed directly starting from the hydrodynamical properties of the model jet (supplied by a check point of the PLUTO code). These quantities, enter the calculation of the emission coefficients, $\epsilon_\nu$, at every computational cell of the 3-D hydrodynamical model grid.

We mention that, the production of synthetic images from the data is carried out by using the line-of-sight code constructed in [12] (here, instead of the radio emission and absorption coefficients we require their $\gamma$-ray equivalents, $\epsilon_\nu$ and $k_\nu$ (in the case of neutrino production we need only emission coefficients).

In using the hydrodynamical code, PLUTO, the energy $E_\gamma$ (in GeV) refers to the observed $\gamma$-ray energy. The quantity $n_{sp}$, i.e. the bulk-flow slow jet proton number density of ejected particles, is the dynamically important. This number density is taken to represent the hydrodynamic number density of the PLUTO code, namely

$$n_{sp} = n_{j(PLUTO)} \tag{14}$$

The fast proton density, $n_{fp}$, even though not-important dynamically, however it is radiatively important. For SS433, in the fast proton power-law energy distribution, the index $\alpha$, takes the value $\alpha$=2, and the ratio of the initial jet beam speed $u_b$ divided by the speed of light is $u_b/c = \beta_b = 0.26$ [13].

In the hydrodynamical (HD) simulations with PLUTO code, as we have done previously [12, 13], the magnetic field lines are assumed to follow the matter flow. Their tangling with the jet material makes applicable the fluid approximation within the jet [33]. In this case, the magnetic field could not affect the flow dynamics which is however possible in the magnetohydrodynamic (MHD) treatment of PLUTO. In the relativistic hydrodynamical version of PLUTO, the magnetic field within the jet's medium is assumed rather strong so as the coupling effects permit the fluid approximation to be applicable. At the same time, the dynamical effects of the magnetic field on the relativistic flow are not permitted [33].

We mention that, in modelling the microquasar SS433 system with PLUTO, only one of the twin jets is considered. The counterjet is presumed to exist outside the model space (at the bottom of $x - z$ plane), but its interference with our model system is considered very small [34]. The computational grid is 3-D Cartesian $(x, y, z)$, homogeneous and the boundary conditions are atopted to be reflective at the jet's base ($x - z$ plane) and outflow at all other planes of the computational domain (box).

The grid spans $120 \times 200 \times 120$ (for x, y, z, respectively), in model length units (equal to $10^{10}$cm) and the resolution used is $300 \times 500 \times 300$ (for x, y, z, respectively). The jet emanates from the middle of the $x - z$ plane, at the point (60, 0, 60)$\times 10^{10}$cm and then advances while precessing around the (60, y, 60) line (parallel to the y-axis). The precession angle for the SS433 jet used ($\delta$=0.2 radians), is slightly smaller than the value of 21 degrees of Ref. [34], in order to allow the use of finer resolution.

The center of the companion star is supposed to be outside of the box, at the point (400, 0, 400) while compact object is situated at the point (60, 0, 60), i.e. at the jet's base. We remind that, because the exact orbital separation in SS433 is not well known, this estimation is within an order of magnitude (the objects are orbiting around their centre-of-mass). We also mention that, we assume that the companion star is not included in the model [12]. However, its wind is included through its density which is taken as decreasing with distance r (as $1/r^2$), away from its centre.

Furthermore, we also include a simplified accretion disk wind through the jet's dynamic interaction with both winds. This means that, our model is less realistic as we approach the companion star and accretion disk locations but the results are reliable in the vicinity of the jet. For a detailed discussion related to important phenomena of the jet's interaction zone with nearby winds, the reader is referred to the Refs. [12, 13, 27, 28] and references therein.

### IV. RESULTS OF SIMULATIONS FOR THE NEW JET MODEL SCENARIOS

In this section, we present the results for two new scenarios of initial conditions (referred to the micro-quasar stellar system SS433) obtained as follows: (i) Hydrodynamical simulation carried out by utilizing as main computational tool the 3-D relativistic hydrocode PLUTO and (ii) Gamma-ray emission synthetic images obtained with the line-of-sight integration.

In scenario C, the jet precesses faster than reality, therefore precession effects are enhanced. In this case, the jet involves artificially accelerated precession, in order to better investigate the effects of precession on the surrounding winds, within the limited time-frame of the model run.

In scenario D, the jet is quite heavier than both winds, in order to consider the possibility of a dense jet beam, containing the estimated jet mass flow of SS433 while remaining more focused and more both narrow. The heavier jet of this scenario, crosses the winds with greater ease. Also the effects of its interaction with the winds appear decreased.

We note that, another characteristic scenario would have been a jet much lighter than both winds, but this would have taken longer simulation time, and practically more difficult.

In both cases, the jet begins to expand into the accretion disk wind, but at a more limited pace, due to the increased density of that wind in the model. As soon as the jet head reaches the stellar wind region, however, the jet's expansion rate increases greatly (especially sideways), in the form of a side shock that accumulates ambient matter. At the same time, the accretion disk matter is expelled outwards, from the vicinity of the jet base, forming a 'ring' around the jet. The accretion disk wind is swept in a prominent way, being denser than the stellar wind, leading to the creation of a halo around the jet base (see below).

The structure develops throughout the model run, therefore suggesting the possibility of its persistence later on, when the jet reaches its lobe in the W50 nebula. This is similar (to a certain extent) in structure to that discussed in Ref. [32]. The above scenarios are applied to the SS433 microquasar as described below.

## A. Description of Runs for Scenarios C and D

### 1. Simulations of Scenario C

Scenario C (medium resolution, see Figs. 1, 2, 3 and 4), has been chosen to cover the case of artificially fast precession of the jet. This way we may investigate the effects of precession on the system observables. The precessing jet sweeps across more of the ambient matter in a given period of time, as compared to an otherwise same but non-precessing jet (run2-scenario), with both jet examples considered to be moving through identical surroundings. Consequently, the effects of the precessing jet on its surrounding environment are, in terms of affected volume, more prominent than when precession is absent. More ambient matter is displaced and part of it ends up being dragged along by the jet, albeit at a pace clearly slower than when precession is slower.

The precessing jet model was therefore run, at an accelerated precession rate, and that showed an enhanced effect of sweeping the accretion disk wind matter from the jet cone. This, in effect, caused the formation of an enhanced, outward moving, 'halo', around the jet base (Figs. 1 and 2). The precessing jet advances, first through the accretion disk wind, and then through the stellar wind, opening its path at an accelerating pace, due to meeting with, progressively, lower resistance, due to the falling density of the winds. The latter originated from both the accretion disk and the companion star. The gradient of the stellar wind, within the computational grid, is, however, not big, due to the increased distance from its origin, the companion star.

The jet precession pronounces the sweeping of the inner, denser part of the accretion disk wind, leading, later on, to the formation, of an expanding, approximately torus-shaped halo, surrounding the jet, whose axis roughly coincides with the axis around which the jet precesses. The torus consists mainly of accretion disk wind matter, forming a loose 'barrel', or hollow cylinder (Figs. 1 and 2) around the jet, having a velocity component (clearly slower than the jet, i.e. sub-relativistic) parallel to the jet's precession axis, and a sideways expansion velocity component as well.

In the scale of the simulation, the companion star has a non-negligible distance from the jet base, therefore in the immediate vicinity of the jet base, it is the -more localized- accretion disk wind that dominates over the stellar wind, in terms of density. It is, therefore, the accretion disk wind matter that is mainly expelled from the incoming jet and rushed outwards, swept over by the precessing jet. This finding might suggest a behaviour that, over a longer –than our model run's– timescale leads to the formation of a 'ruff' of material, around the cone swept by the precessing jet, perhaps along the lines of the 'bow-tie' structure recently observed in SS433 [32].

The jet precession makes possible dragging an increased quantity of surrounding matter, along the jet, compared to a non-precessing jet, where the swept matter is significantly less. This happens because the precessing jet covers a cone with an opening angle much larger than the jet's own. Therefore, more ambient matter is displaced than from a straight jet. Furthermore, the partially sideways motion of the precessing jet further disturbs the surrounding winds, pushing and dragging them in an outwards direction. However, the time scale for precession is much bigger than the jet crossing time of the model space. Therefore, a longer term simulation, perhaps including replenishment of the winds as well, would be needed in order to study the effects of precession on the jet's environment.

### 2. Simulations of Scenario D

In the Scenario D (medium resolution) discussed in this work (see Figs. 5, 6, 7 and 8), the jet is assumed heavier than its surrounding winds, which in turn are also somewhat heavier than those of the other cases, leading to a faster crossing of the computational domain (Fig. 5). Sideways expansion is also swift, as the increased jet-mass density allows for a faster 'sweep' of the wind matter. Leftovers from the displaced accretion disk wind matter can be seen piling up around the jet base (Fig. 6).

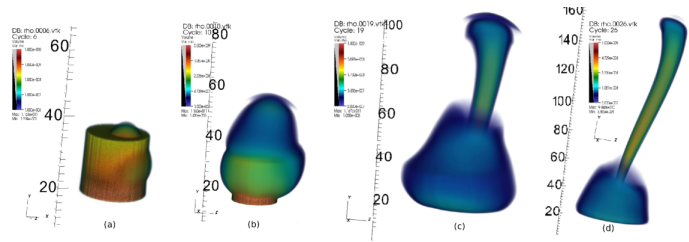

FIG. 1. Scenario C: Density snapshots obtained as in scenario A of Ref. [13], but now we added an accelerated precession of the jet that leads to the formation of a revolving jet flow. See the rightmost panel of Table I, where the jet is the most evolved among the plots (parameters of run3, snapshot intervals 50/count). The jet's behaviour resembles to that of Scenario A (run1) of Ref. [13]. Also, now the jet sweeps a larger volume of ambient matter and propagates slower than in the non-precession case, though now more ambient matter is further activated for emission. Matter piles up around the jet base, hinting on a halo of slower moving (outbound) material there, crudely reminiscent of the suggestion of [32] for a halo formed around SS433.

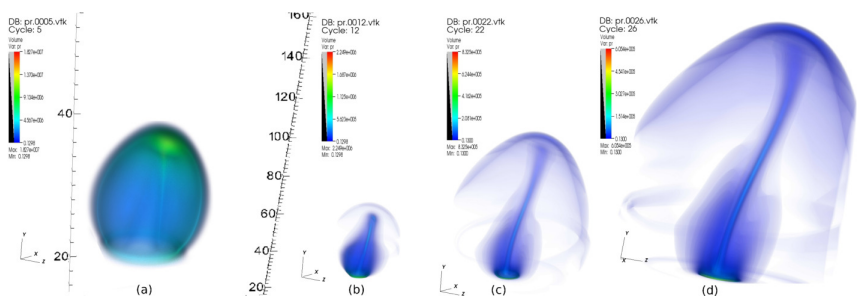

FIG. 2. Scenario C: 3-D illustrations of pressure evolution in the vicinity of the jet, as well as in the jet itself (in this case linear plots are produced, in order to better display the periphery of the system (see run3 parameters in Table I). The three rightmost plots (b, c, d) share the same spatial size scale, while (a) has been magnified. The jet precession is visible, especially in (c) and (d), while a pressurized mass concentration can be found in the vicinity of the jet base, persisting throughout the model run and slowly advancing outwards. The front of the jet itself, due to precession, advances in an asymmetrical way.

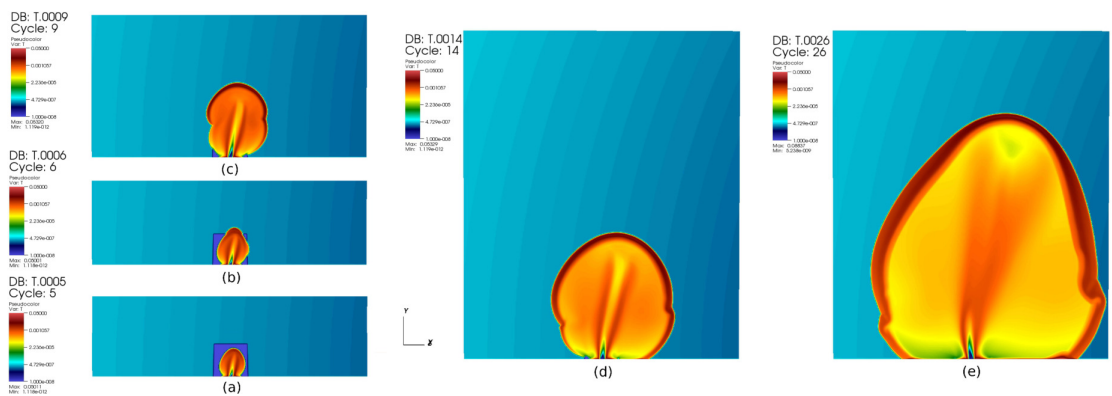

FIG. 3. Scenario C: A series of slices representing the time evolution of temperature, as in scenario A of [13], but now assuming increased precession rate (see text). The precessing jet can be seen (d,e) to sweep across a larger (compared to scenario A [13]) portion of the surrounding wind volume, piling up additional matter around its head and sides (see run3 parameters in Table I). The expansion rate is lower in the accretion disk wind region and increases in the stellar wind region, but afterwards it stays relatively constant, as the density gradient of the wind (visible in the images) is not too large along the jet path.

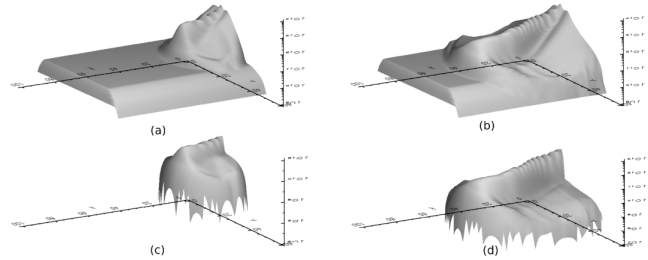

FIG. 4. Scenario C (run3): Illustration of the synthetic $\gamma$-ray images. In the plots the x-y plane defines the (synthetic) observation plane. $z$-axis represents the intensity at each pixel of the observation plane (arbitrary units). For each one of the latter, a line of sight (LOS) is drawn that crosses the computational domain volume and ends up at the aforementioned pixel. The radiative intensity along the LOS, is calculated using the radiative transfer equation. We see two different cases for emission coefficient used, with two snapshots for each case. In the first row, $\rho$ is used as emission coefficient (the intensity appears higher when the matter is denser). In the second row, $\rho u^2$ is used as emission coefficient. The fast moving jet matter now prevails in terms of $\gamma$-ray intensity, as compared to the clearly lower emission from slower moving surrounding material. In both configurations, we see the distinct signature of the precessing jet on the bent jet emission patterns formed, as well as on resolution effects at a faster pace than the rest of runs.

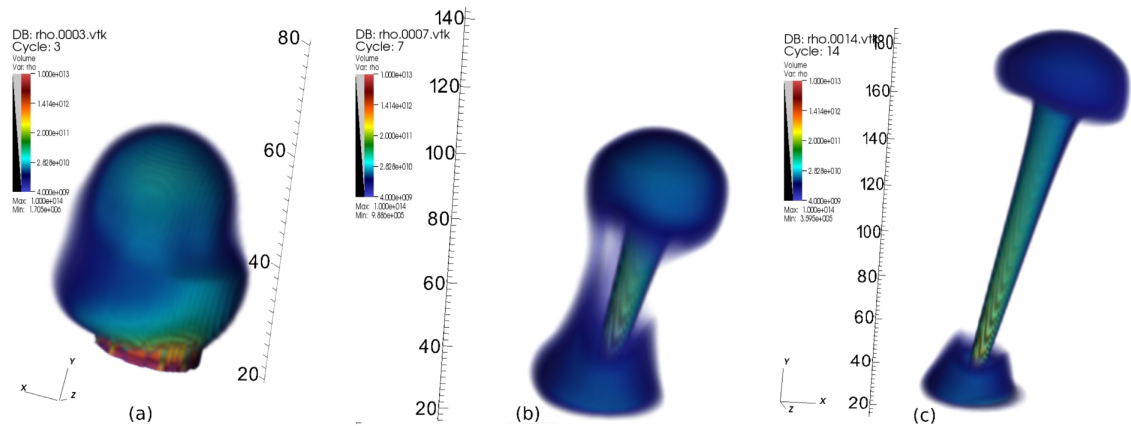

FIG. 5. Scenario D: A series of snapshots depicting the density evolution (logarithmic plots) for the run4. This time the jet is quite much heavier than both surrounding winds, leading to a decreased jet crossing time of the model space. The jet mass flow rate now rests closer to the estimates for SS433. The jet breaks through the accretion disk wind construct and soon crosses the stellar wind at a rapid pace (run4 parameters in Table I, snapshot intervals of 50/count).

The jet forms a funnel that transfers mass outwards at an increased flow rate. The properties of the jet's surroundings are now less pronounced, as the expansion meets with reduced resistance from ambient matter. The dynamic behaviour of the jet dominates the hydro simulation, with the wind's matter giving way to the jet (Figs. 7 and 8). This case covers the possibility of jet production as a very dense inflow at the source, thus enriching nearby interstellar matter with important mass outflow per time interval.

All of the hydrocode snapshots, of hydrodynamic parameters, have been created with the VisIt visualization code, whereas the synthetic $\gamma$-ray images have been produced using IDL.

### B. Gamma-ray emission synthetic images

The above discussed runs with the hydrocode PLUTO have been performed for a precessing jet model of SS433. As the simulation proceeds, at some point the computational model space data is transferred to an output file to be processed (with the available line of sight (LOS) code) for producing a synthetic $\gamma$-ray image of the system. The data of a snapshot from the PLUTO hydrocode is transferred, in the form of 3-D data arrays (density, velocity vector, Pressure: $\rho, u_x, u_y, u_z, P$) to a routine that performs the LOS integration [12, 13]. Along the LOS, $\gamma$-ray emission and absorption coefficients are provided at each point.

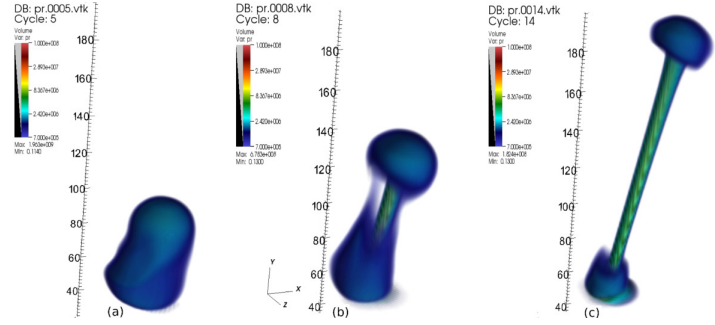

FIG. 6. Scenario D: The pressure evolution for the model jet of run4. The patterns are similar to those of the density for the same run. The higher jet's density dominates the system dynamics, leading to comparatively poorer features around the jet, in relation to other runs where the jet was lighter. The pressure distribution follows the above general pattern. A pressure increase is also seen around the jet base, due to the mass leftover from the disturbed accretion disk wind that used to be there (run4 parameters in Table I).

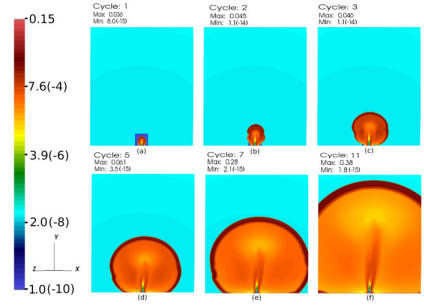

FIG. 7. Scenario D: A series of plots for the jet temperature of run4 (run4 parameters in Table I). The, faster than previous runs, advance of the jet, through surrounding wind matter, is characterized by an expanding shock front. Said expansion is initially slower, till the jet crosses the simplified accretion disk wind construct. Then, a higher rate of expansion occurs, especially to the sides, as the stellar wind gradient is less pronounced, since we have: $\Delta x < \Delta r$, where $x$ is the jet crossing distance and $r$ is the distance to the binary companion star.

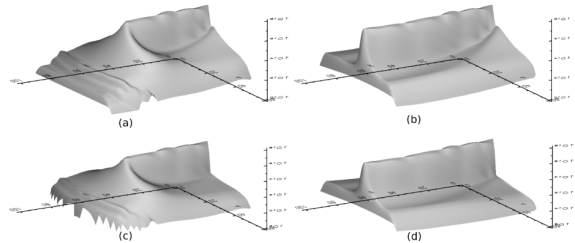

FIG. 8. Scenario D: The top row shows two snapshots of the 'heavy jet' run4 (run4 parameters in Table I), using the hydrodynamical density $\rho$ as the emission coefficient (arbitrary units). The jet makes its way through the surrounding winds, piling up matter ahead of its head and around the jet shock front, as well as around the jet base (remnants of the accretion disk wind). The bottom row shows the line of sight integration images of the same scenario (D), where now $\rho u^2$ is the emission coefficient. This time only matter that is both fast and dense, simultaneously, does contribute to the emission. The jet base, therefore, now emits much less, since, only the jet flow there, is both fast and dense, whereas the remnants of the wind constitute dense but slow matter. The impact of resolution effects, from near the jet base, can also be seen as humps on the back of the jet intensity ridge.

At this level, the orbital separation of the compact object and the donor star is taken to be about $4\times10^{12}$cm, so the stellar wind origin, does not coincide with the jet base. Also, the jet is taken to travel through a halo produced from the accretion disk (centered at the jet's base). This matter is commonly called the accretion disk wind) [12, 13].

With our method we follow two separate steps of calculations. We first obtain the hydrodynamic quantities of density and velocity with the PLUTO hydrocode and second, the integration (with the LOS codes) provides the intensity. This de-coupling allows, in the calculation of emission, the LOS integration to be performed using either $\rho$ (in CGS units) or $\rho v^2$ (in speed of light dimensionless units, $c = 1$), as the emission coefficient. Then, the synthetic image is formed, and subsequently the values of all of its pixels are added up to provide the total intensity released from the studied object. Finally, the $\gamma$-ray emission calculation is performed, separately in Mathematica (for unit proton number density).

In each of the above scenarios (runs), for $\gamma$-rays a synthetic image was produced, at a suitable model time, using the relevant radiative transfer code of [12] (only the emission coefficient was employed). As can be seen, in all the synthetic $\gamma$-ray images (Figs. 4 and 8), at each point of the computational volume, the denser the matter, the higher the emission at that point is.

Furthermore, it is clear that the bigger the number of significant emission points along a line of sight, the higher the total emission of the whole line of sight is. By adding the dependence on the local velocity, denser but slower matter cannot emit any significant amount of $\gamma$-rays. Therefore, emission from the jet body and its interaction zone with surrounding media can be seen to be stronger in the $\rho\beta^2$ maps. In addition, the rest of the (roughly inert) medium in the system contributes very little to $\gamma$-ray emission.

Before closing we note that, currently we apply the above mentioned method to carry out jet emissions simulations for other microqasar systems like the Cygnus X-1 and Cygnus X-3. From the viewpoint of observations we should mention that, for lower energy $\gamma$-rays, orbital platforms, such as NASA's Fermi and ESA's INTEGRAL, already offered an important relevant body of observations for various systems [20, 35]. Very high energy $\gamma$-rays can also be studied using data provided by ground based Cerenkov telescopes such as HESS, MAGIC and HEGRA. Hydrodynamical jet models, combined with artificial imaging, offer realistic estimates of the conditions in the jet and surrounding environments, adding insight to open questions about the $\gamma$-ray jet emission from a microquasar.

## V. SUMMARY AND CONCLUSIONS

A precessing jet was modelled using the relativistic hydrodynamic code (PLUTO). Furthermore, the results were processed using the line of sight code, assuming that the flow velocity $u$ is much smaller than the speed of light. The LOS code integrates along lines of sight the equation of radiative transfer, without scattering. The emission coefficients are, in general, a function of hydrodynamical and also of radiative parameters.

The intensity result of each LOS is assigned to the pixel where the LOS meets the imaging plane of 'observation'. This way, an image is formed, which could be called a synthetic $\gamma$-ray image. Only emission is used for this paper, but absorption may both be incorporated.

The currently available resolution for $\gamma$-ray is lower than the resolution of synthetic imaging. Yet the connection between the emission properties and the underlying system dynamics do offer useful constrains on a variety of system parameters, such as the jet kinetic luminosity energy $L_k$. The above occur in the light of potential future observations, originating from orbital $\gamma$-ray telescopes, from terrestrial Cerenkov detector arrays and even from underground neutrino detectors.

## ACKNOWLEDGEMENTS

Dr. Odysseas Kosmas wishes to acknowledge the support of EPSRC via grand EP/N026136/1 "Geometric Mechanics of Solids".

# CONFLICTS OF INTEREST

The authors declare that they have no conflicts of interest.

---